\documentclass[preprint,12pt]{elsarticle}

\usepackage{amssymb}
\usepackage[utf8]{inputenc}
\usepackage{graphicx}
\usepackage{amsmath,amssymb,amsfonts}
\usepackage{algorithm}
\usepackage{multirow}
\usepackage{algorithmic}
\usepackage{changepage}
\usepackage{textcomp}
\usepackage{subcaption}
\usepackage[table]{xcolor}

\journal{Journal of Network and Computer Applications}

\begin{document}

\begin{frontmatter}

\title{PTTS: Zero-Knowledge Proof-based Private Token Transfer System on Ethereum Blockchain and its Network Flow Based Balance Range Privacy Attack Analysis}

\author[inst1]{Goshgar Ismayilov}

\author[inst1]{Can Özturan}

\affiliation[inst1]{organization={Department of Computer Engineering},
            addressline={Bogazici University}, 
            state={Istanbul},
            country={Turkey}}

\begin{abstract}
Blockchains are decentralized and immutable databases that are shared among the nodes of the network. Although blockchains have attracted a great scale of attention in the recent years by disrupting the traditional financial systems, the transaction privacy is still a challenging issue that needs to be addressed and analysed. We propose a \textit{P}rivate \textit{T}oken \textit{T}ransfer \textit{S}ystem (PTTS) for the Ethereum public blockchain in the first part of this paper. For the proposed framework, zero-knowledge based protocol has been designed using Zokrates and integrated into our private token smart contract. With the help of web user interface designed, the end users can interact with the smart contract without any third-party setup. In the second part of the paper, we provide security and privacy analysis including the replay attack and the balance range privacy attack which has been modelled as a network flow problem. It is shown that in case some balance ranges are deliberately leaked out to particular organizations  or adversial entities, it is possible to extract meaningful information about the user balances by employing minimum cost flow network algorithms that have polynomial complexity. The experimental study reports the Ethereum gas consumption and proof generation times for the proposed framework. It also reports network solution times and goodness rates for a subset of addresses under the balance range privacy attack with respect to number of addresses, number of transactions and ratio of leaked transfer transaction amounts.
\end{abstract}


\end{frontmatter}

\section{Introduction}

Blockchain is basically a decentralized ledger of transactions that is publicly shared among the nodes in the network. It was first proposed by the seminal paper of Bitcoin in 2008 \cite{bitcoin}. Within these fourteen years, it has already revolutionized the traditional financial systems worldwide \cite{bellavitis2020}. Blockchain technology provides decentralization without relying on any trusted third-party, data integrity, immutability, traceability and auditability \cite{buterin}. These benefits have been intensively exploited to solve many real-world problems including biometrics \cite{mohatar2019}, e-voting \cite{koc2018, fatrah2019} and bartering \cite{ozturan2020}. However, despite of all these significant advancements, blockchain has still numerous challenges and issues that need to be thoroughly addressed including privacy over transactions, security, resource waste, interoperability, throughput and scalability \cite{yli2016}. Especially, privacy over transactions is about to what extent the transaction data should be publicly revealed in the network. 

This private transaction issue in the blockchain technologies has been already addressed in the literature. The proposed systems have mainly followed two different approaches as extending the architecture of already in-service blockhain or developing completely new privacy-preserving blockchain. The popular systems under the first approach include Tornado Cash \cite{tornadocash} and Aztec \cite{aztec} for the Ethereum blockchain while the popular systems for the second approach include Zerocash \cite{zerocash} and Zcash \cite{zcash}. These two approaches have several pros and cons over the others where the second approach inherently provides more flexibility and customizability while it must consider the other aspects of blockhain from the scratch. In this paper, we have followed the first approach where our private token transfer system has been integrated to the Ethereum blockchain through a smart contract.

The advancements in the blockchain technologies during the past decade have led several cryptographically strong and privacy-preserving zero-knowledge proof protocols to be developed. The popularly used ones among these protocols include zk-SNARKs \cite{zksnark} proposed in 2014, zk-STARKs \cite{zkstark} proposed in 2018, Bulletproofs \cite{bulletproofs} proposed in 2018 and Plonk \cite{plonk} proposed in 2019. Up to today, the protocols have been successfully implemented for plenty of different problems in blockchain including sealed-bid auctions \cite{li2021}, traffic managements \cite{li2020, meese2020, gudymenko2020}, internet-of-things \cite{soewito2021, singh2020} and energy systems \cite{pop2020}. The common driving motivation for these applications of zero-knowledge proof protocols into the blockchain technologies is mainly originated from information asymmetry principle where certain actors available in the network possess more information than the other actors. In this paper, we have applied zk-SNARKs protocol through Zokrates privacy-preserving tool \cite{zokrates} to the Ethereum blockchain in order to develop our private token transfer system.

It has been observed in the literature that the strong cryptographic schemes and protocols are assumed to be sufficient to protect the private token transfer systems from the external vulnerabilities without considering the benefits of the rational system users. However, it may be  possible in certain circumstances that a certain portion of the private blockchain 
transactions may be deliberately or mistakenly leaked out to the specific malicious entities or non-malicious organizations
(e.g. exchange services, governmental units). Such organizations may aim to gradually collect the private transactions from the users in return of certain benefits including money. The more the amount of the private transactions are collected, the clearer global transaction graph is and consequently the more precise the user balances can be estimated. In the most ideal case where all the private transactions are collected, the user balances can be directly found without any need for the estimation. We are not aware of any work to address this issue in the literature. Therefore, we have modelled the balance range privacy attack as a minimum cost flow network problem
in this paper. We assume that a subset of the transactions  are leaked out to a particular entity. The solutions of the minimum cost flow network problems
help us to estimate the possible ranges for the user balances. The validity of our 
approach is also supported through the experiments in our work. 

The main contributions of this paper can be listed as follows: 

\begin{itemize}
\item A \textit{P}rivate \textit{T}oken \textit{T}ransfer \textit{S}ystem (PTTS) is proposed for Ethereum, for which a zero-knowledge proof protocol, a private token smart contract and a web user interface are designed and developed.

\item Security analysis of the PTTS framework is made by considering the replay and the balance range privacy attacks.

\item A novel modelling of the balance range privacy attack is introduced as a minimum cost flow network problem where it is shown that meaningful information about user balances can be extracted in case some transactions are leaked.

\item Extensive experiments are performed where gas consumption and proof generation times are considered for the proposed framework while network solution times and goodness rates 
(defined in Subsection~\ref{experimentsbalanceattack}) with respect to small number of addresses are considered for the balance range privacy attack.
\end{itemize}

The remainder of the paper is organized as follows. The Section 2 reviews the necessary background and related work. Section 3 presents the private token transfer system including zero-knowledge proof design, smart contract design and web user interface. Section 4 analyses the proposed framework with respect to the replay attack and the balance range privacy attack for which a minimum cost flow network design is presented. Section 5 discusses the experimental results for the private token transfer system and the attacks as well. Finally, Section 6 concludes the paper. 

\section{Background and Related Work}
We have provided background and related work on the topic of zero knowledge proofs 
in the following subsections. For the background work on the topic of minimum cost network flows, the books \cite{gutin2008, ahuja1988} can be consulted. 

\subsection{Zero-Knowledge Proof}

Zero-knowledge proof was first proposed by Goldwasser, Micali and Rackoff in 1985 \cite{zkp}. It basically refers to a group of cryptographic protocols which allow an actor (i.e. prover) to convince another actor (i.e. verifier) about the validity of a particular statement without disclosing the statement itself. The prover is responsible for asserting a statement beforehand and generating a corresponding proof for that while the verifier is responsible for validating the correctness of the given proof. There are two main elements that differ zero-knowledge proof from traditional mathematical proof as interaction and randomization \cite{blum1991}. The interaction requires the prover and the verifier to communicate back and forth while randomization requires several randomized information to be exchanged where there exist negligible probability that the prover may convince the verifier about a false statement.

The zero-knowledge interactive proof protocols must satisfy the following three properties \cite{zkp}: 

\begin{itemize}
\item \textit{Completeness}: Any true statement must have proof to convince the verifier. In other words, the prover can successfully construct a proof for a true statement that the honest verifier will eventually accept.

\item \textit{Soundness}: No false statement must have proof to convince the verifier. No dishonest prover can construct a proof for a false statement such that the honest verifier will eventually accept. However, there still exist a very small probability that the dishonest prover can cheat. 

\item \textit{Zero-knowledge}: If the proof is correct, the verifier cannot learn anything more than the validity of the statement. For this property, there must exist a simulator algorithm through which any knowledge learned during the interaction can also be learned by the verifier itself without any interaction with the prover.

\end{itemize} 

Based upon these properties, a proof is complete if the first property is satisfied; sound if the second property is satisfied; and zero-knowledge if the third property is satisfied. The first two properties are the general characteristics of the interactive proof protocols. However, it is the third property that differs the zero-knowledge proofs from the other interactive proofs. The zero-knowledge proofs can be also classified with several ways as (1) proof and argument systems; (2) perfect, statistical and computational systems; and (3) interactive and non-interactive systems \cite{nguyen2006}. However, it is possible to convert the interactive zero-knowledge proofs into non-interactive zero-knowledge proofs through Fiat-Shamir Heuristic \cite{fiatshamir}.

The zero-knowledge proof protocols in the literature can be generated based on different approaches, mainly as number theoretic and graph theoretic approaches. The most popular protocols for the former category include Schnorr's Protocol \cite{schnorr} based on discrete logarithm, Feige-Fiat-Shamir Protocol \cite{feigefiatshamir} based on integer factorization and Guillou-Quisquater’s Protocol \cite{quisquater1988} based on integer factorization. On the other hand, the most popular protocols for the latter category include graph-3 colorability problem, graph isomorphism, graph non-isomorphism from NP problems \cite{goldreich1991}. Besides from those traditional protocols, there are several recently proposed complex protocols as well including zk-SNARKs \cite{zksnark}, Bulletproofs \cite{bulletproofs} and Plonk \cite{plonk}. Although they have relative strengths and weaknesses towards the others, it can be said that zk-SNARKs perform better in terms of proof size and time complexity \cite{ortiz2020}.

\subsection{Zokrates}

Zokrates is a scalable privacy-preserving toolbox developed for the Ethereum blockchain to provide verifiable computations \cite{zokrates}. It is based on the specific family of zero-knowledge protocol as zk-SNARKS (Zero-Knowledge Succinct Non-Interactive Arguments of Knowledge) \cite{zksnark}. The most significant attributes of such proofs are that they are (1) succinct in terms of short proof length and fast proof verification, (2) non-interactive where no interaction is required between prover and sender and (3) zero-knowledge where the three main properties of the zero-knowledge proof are satisfied. 

There are two main motivations behind the proposal of Zokrates \cite{zokrates}. The first motivation is that Ethereum significantly limits to perform the complex mathematical operations on-chain by requiring high gas cost. Therefore, Zokrates reduces the on-chain computational effort to a great extent by allowing the proofs to be generated off-chain and to be verified on-chain without disclosing any secret information. The second motivation is that the implementation and the integration of zero-knowledge proof to blockchain is hard and error-prone. For this reason, Zokrates proposes a domain-specific language over which the necessary verifiable computations can be implemented with high-level of abstraction. 

The architecture of the Zokrates framework consists of six different inter-connected modules as parser, flattener, witness generator, R1CS converter, libsnark and contract generator \cite{zokrates}. The compiler has the first two modules as parser and flattener where it transforms the high-level code into the flattened code. Later, the witness generator module collects the flattened code as well as user private and public inputs to generate a witness which will be used during proof generation. The R1CS module converts the flattened code into Rank-1 Constraint System in order to feed the libsnark module which is a cryptographic library implementing zk-SNARKs in C++ programming language \cite{libsnark}. The libsnark module generates a proving and a verifying key where the proving key is public and used during off-chain proof generation while the verifying key is also public and used during on-chain proof verification \cite{ballesteros}. After the witness and the keys are ready to be used, the proof is generated. In the final stage, the contract generator generates a smart contract that will be deployed to the blockchain in order to verify the proof using the embedded verifying key. 

\subsection{Private Transaction in Blockchain}

There exist several private transaction approaches already proposed in the literature. The first approach is Tornado Cash \cite{tornadocash} where it is based on a fully decentralized smart contract accepting tokens from source addresses so that they can be later withdrawn by destination addresses. The main purpose of the system is to break the on-chain link between the source and destination addresses in order to prevent address linkability and traceability. Based on a special version of zero-knowledge proof protocol called zk-SNARKs \cite{zksnark}, Tornado Cash is especially developed for ERC-20 token transfer operations on Ethereum. However, it supports several other networks as well including Avalanche. Any user wishing to involve with private transaction through Tornado Cash must generate a randomly generated key and deposits a certain amount of tokens to the smart contract along with the hash of the random key. Later, the same amount of tokens can be withdrawn as long as the valid proof corresponding to the random key is provided. 

Another private token transfer system has been built by the Kaleido company, especially for ERC-20 tokens using zero-knowledge proof \cite{kaleido}. It offers a special service to the users with private balances and private identities where the zero-knowledge proof protocol is based upon the extended version of Zether private payment mechanism \cite{zether}. Although the privacy of balances are protected through zero-knowledge proof where transaction amounts are replaced with random bytes, the privacy of identities are provided by decoy addresses to multiplex the number of source and destination addresses. The private token transfer operations are performed on three-staged flow as fund-transfer-withdraw. In the first stage, certain amount of ERC-20 tokens are deposited to the smart contract in return of private ZTH tokens. The users can make private transfers through these tokens. Later, the ZTH tokens can be exchanged back to the original ERC-20 tokens. 

Zerocash as another privacy-preserving token transfer system extends the Bitcoin blockchain with an extra layer of zero-knowledge protocol \cite{zerocash}. As in Tornado Cash, the version of zero-knowledge protocol used in this system is zk-SNARKs as well \cite{zksnark}. The privacy token transfer design behind Zerocash is similar to the design of Kaleido where there exist two types of exchangeable coins in Zerocash as \textit{basecoins} and \textit{zerocoins}. While the transaction data is totally public in basecoins, it is privatized in the zerocoins including source and destination addresses and transaction amount. This can be interpreted  as the users can anonymously deposit and withdraw as long as they use zerocoins. The \textit{mint} transactions are used to convert basecoins into zerocoins where commitment operations are performed to map the public coin value into the private coin value. For the \textit{pour} transactions, these commitments are used in the zero-knowledge proof protocol so that the users can validate the correctness of the private coin value claimed. 

Zcash is an individual blockchain network that supports privacy-preserving transactions using zero-knowledge proof \cite{zcash}. As in Tornado Cash and Zerocash, the zero-knowledge proof protocol in Zcash is also based on zk-SNARKs. Different from the previous systems, Zcash has two different address types as z-addresses which are private and t-addresses which are public. Based on these two address types, there are four different transaction types as private transaction between z-to-z addresses, deshielding transaction between z-to-t addresses, shielding transaction between t-to-z addresses and finally public transaction between t-to-t addresses. The source and destination addresses along with the transaction amount are protected in the private transactions. Another novelty of Zcash is that it supports viewing keys so that third parties can read the transaction data without any write access as long as viewing key is provided for the sake of governmental auditability. 

AZTEC, standing for Anonymous Zero-Knowledge Transactions with Effcient Communication, is a zero-knowledge protocol which is built on the top of Ethereum and managed through a smart contract called as AZTEC Cryptographic Engine \cite{aztec}. It basically enables privacy on public blockchain to allow users to create, send, swap and pay confidential tokens. The difference between the previously mentioned systems and AZTEC is that AZTEC is primarily designed to solve the interoperability between tokens of different smart contracts. Although the mechanism of AZTEC is relied on zero-knowledge proofs and homomorphic encryption to exchange the tokens into special AZTEC notes, the resulting transactions in the blockchain are still verifiable by the validator nodes. The common functionalities supported by AZTEC include (1) join-split, (2) bilateral swap, (3) dividend proof, (4) mint, (5) burn, (6) private range and (7) public range. The private range function is used to prove that an AZTEC note is greater than another note while the public range function is used to prove that an AZTEC note is greater than a public integer.

\section{ Private Token Transfer System (PTTS) }

\subsection{Problem Definition}

\begin{figure*}[htbp]
\centerline{\fbox{\includegraphics[width=0.9\textwidth]{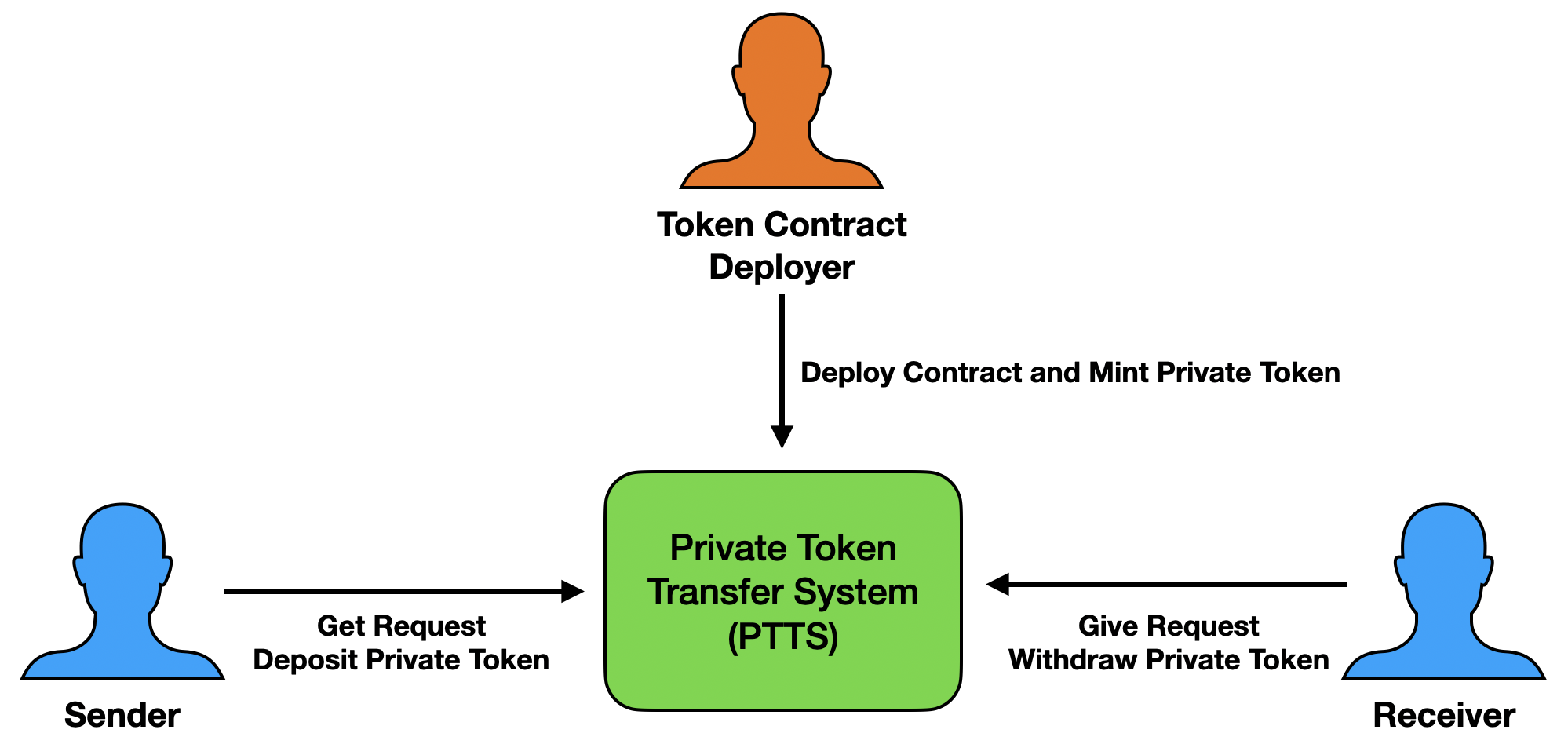}}}
\caption{The Sequence Diagram For the Proposed Private Token Transfer System}
\end{figure*}

Information asymmetry refers to an imbalance between two parties in interaction, in which one party possesses more knowledge than the other one \cite{ortiz2020, akerlof1978}. In blockchain technologies, there currently exists certain amount of information asymmetry where the ordinary users having transaction data need privacy during token transfer while the validators in the network requires that transaction data to be transparent for accurate validation and auditability. The extent of privacy to be evaluated during transaction may change based on the problem definition itself. Certain problems may consider shielding only transaction amounts and user balances \cite{aztec} while the others may also consider shielding user addresses \cite{zcash}. Our private transaction problem is based on this conflict where the transaction amounts and the user balances must be private but validatable while the privacy of user addresses is out of the scope. For this problem, we have asked the following question: "Is it possible to transfer tokens privately on public blockchain without any third-party setup?".

Our private transaction problem consists of five consecutive stages as (1) deploying token contract, (2) getting consent for token deposit, (3) giving consent for token deposit, (4) depositing tokens and (5) withdrawing tokens. These stages are also shown in the Fig 1. where there exists three main actors as token contract deployer, token sender and token receiver. The token owner is responsible to deploy its own private token to blockchain in the first stage. Later, the main responsibilities of the sender are to (1) get consent from the receiver at the second stage since not all the users may wish to be involved with private transaction and (2) deposit tokens privately to the smart contract in case consent is given by generating corresponding proof so that the transaction can be validated in the fourth stage. On the other hand, the main responsibilities of the receiver are to (1) give consent if there already exists a request pending in the third stage and (2) withdraw those tokens privately from the smart contract by generating corresponding proof for the transaction validation in the fifth stage. 

The private transaction problem has several issues to be resolved during solution architecture design. They mainly include (1) privacy, (2) public verifiability, (3) authentication and (4) non-interactivity. The first  privacy issue requires the system to hide transaction amounts and user balances. Transfer amounts must be known by both the sender and the receiver while the balances must be known by users themselves. The second issue as public verifiability requires the system to allow the transactions to be validated even if certain transaction data is still private. The third issue as authentication must allow no unauthorized sender to deposit private tokens and similarly, no unauthorized receiver to withdraw private tokens. Finally, in the fourth issue as non-interactivity, the sender and the receiver of the transaction must interact with each other as little as possible for the fast and efficient private token transfer.

\subsection{Zero-Knowledge Proof Design}

In our private token transfer system, the direct transaction validation approach in the blockchain network through the public data has been replaced with the indirect validation approach using our zero-knowledge proof protocol. Our PTTS protocol has been developed based on the following  initial ideas and Zokrates 
implementations by~\cite{lundkvist,github1, github2}. In the developed system, the SHA-256 cryptographic commitment scheme is used to bind each private data with a public commitment. Due to the binding property of the scheme, it is not possible to associate the commitment with another private data. Similarly, due to the hiding property of the scheme, it is also not possible to extract the public data from the commitment itself. The commitments corresponding to the user balances are stored in our private token smart contract. It means that whenever the users wishes to deposit or withdraw certain amount of tokens, they have to generate the proofs based on their claims and the commitments. If the bindings between the claims and the commitments are correctly built, the proofs are generated off-chain and later successfully validated on-chain in the smart contract. 

Our zero-knowledge proof protocol including one sub-protocol for senders and another sub-protocol for receivers has been implemented on Zokrates domain-specific language. Their pseudocodes are presented in Algorithm 1 and Algorithm 2, respectively for senders and receivers. For the first pseudocode, the current balance of the user and the transaction amount deposited to the smart contract are provided as private fields which are known only by sender itself. The next user balance can be easily computable with respect to the current balance and the transaction amount. The public commitments corresponding to the current balance, the next balance, the transaction amount are also provided. Later, Zokrates checks four different conditions to deposit the tokens as: (1) the transaction amount to be transferred must be less than the current balance, (2, 3, 4) the commitments provided for the transaction amount, the current balance and the next balance must be equal to the computed commitments. Since the commitments for the current balance are stored in the smart contract, the senders cannot cheat on their claims. If they were not stored, the senders would be able to generate proof for any balances they wish.

It should be noted that three different blind factors (for the current balance, the next balance and the transaction amounts) are also used during the zero-knowledge proof generation in our protocol. The primary reason behind the blind factors is to prevent preimage attack. Suppose that no blind factors were used, then an attacker in such a case would be able to use enumeration or look-up table to find the commitments for the possible balance and transaction amount values. However, since the blind factors are only known by the sender itself in the current design, it is not possible to perform preimage attack as long as the blind factors are large numbers. Based upon all the architectural details provided above, it can be said that the resulting verifiable proof that Zokrates generates off-chain is later sent to our private token smart contract. In case it is successfully validated on-chain, the amount of tokens once given in the user claim are deposited to the smart contract and user balance is updated. 

\begin{algorithm*} 
\small
\caption{Zokrates Code for Sender Proof}  
\begin{algorithmic}[1]

\STATE \textbf{def} main(\textbf{private} amount, \textbf{private} balance, \textbf{private} amountBlindFactor, \textbf{private} balanceBlindFactor, \textbf{private} nextBalanceBlindFactor, \textbf{public} amountHash, \textbf{public} balanceHash, \textbf{public} nextBalanceHash): 

\STATE \;\; computedAmountHash = \textbf{sha256}(amountBlindFactor, amount)

\STATE \;\; computedbalanceHash = \textbf{sha256}(balanceBlindFactor, balance)

\STATE \;\; computedNextBalanceHash = \textbf{sha256}(nextBalanceBlindFactor, balance - amount)

\STATE \;\; result = \textbf{if}(amount \textbf{$\leq$} balance \textbf{\&\&} computedAmountHash \textbf{==} amountHash \textbf{\&\&} computedbalanceHash \textbf{==} balanceHash \textbf{\&\&} computedNextBalanceHash \textbf{==} nextBalanceHash) \textbf{then} 1 \textbf{else} 0 \textbf{fi}

\STATE \;\; \textbf{return} result

\end{algorithmic}
\small
\end{algorithm*}

\begin{algorithm*} 
\small
\caption{Zokrates Code for Receiver Proof}  
\begin{algorithmic}[1]
\STATE \textbf{def} main(\textbf{private} amount, \textbf{private} balance, \textbf{private} amountBlindFactor, \textbf{private} balanceBlindFactor, \textbf{private} nextBalanceBlindFactor, \textbf{public} amountHash, \textbf{public} balanceHash, \textbf{public} nextBalanceHash): 

\STATE \;\; computedAmountHash = \textbf{sha256}(amountBlindFactor, amount)

\STATE \;\; computedbalanceHash = \textbf{sha256}(balanceBlindFactor, balance)

\STATE \;\; computedNextBalanceHash = \textbf{sha256}(nextBalanceBlindFactor, balance + amount)

\STATE \;\; result = \textbf{if}(computedAmountHash \textbf{==} amountHash \textbf{\&\&} computedbalanceHash \textbf{==} balanceHash \textbf{\&\&} computedNextBalanceHash \textbf{==} nextBalanceHash) \textbf{then} 1 \textbf{else} 0 \textbf{fi}

\STATE \;\; \textbf{return} result
\end{algorithmic} 
\small
\end{algorithm*}

There are some differences between the first pseudocode for senders and the second psuedocode for receivers. The first difference is that the first condition of checking whether the transaction amount is smaller than the current balance or not is not necessary any more for the receiver. The second difference is that the transaction amount must be subtracted from the sender balance while it must be added into the receiver balance. It should be also noted that the receiver must know certain private data to construct the proof as transaction amount and corresponding blind factor, which means that they should be conveyed from the sender to receiver through a secure channel. This channel is realized through elliptic curve cryptography based asymmetric encryption scheme where the sender uses the public key of the receiver in order to encrypt the private data while the receiver uses its own private key in order to decrypt the private data. Thanks to the web user interface discussed later, the system users should not carry any concern about the complex zero-knowledge proof, cryptographic commitment and asymmetric encryption operations.

\subsection{Smart Contract Design}

The private token smart contract provides several necessary functionalities to orchestrate the possible operations between the senders and the receivers, based on  based on the four-staged private token transfer system. From the perspective of the smart contract, each sender has two core responsibilities as getting consent from the receiver and later depositing tokens privately if consent is given; and each receiver has also two core responsibilities as giving consent to the sender and later withdrawing tokens privately. Our private token smart stores the hashes of the senders and the receivers privately by encoding with the corresponding blind factors. 

The function for the first stage of the framework which is \textit{getting consent} is given in Algorithm 3. It requires the receiver address as an input to which the sender will send tokens. To manage the requests between senders and receivers, the smart contract uses a global mapping variable named as \textit{request} which is boolean type. The keys of the mapping are the sender and the receiver addresses while the values of the mapping can be either true or false. Intuitively, the value of true means that there is a request from the sender to the receiver to send tokens and vice versa. Once the function is called, the value for the correct keys in the mapping is simply set to true.

\begin{algorithm} 
\small
\caption{Sender Getting Consent}  
\begin{algorithmic}[1]
\STATE \textbf{function} getConsent(\textbf{address} \_to) \textbf{public} \{
\STATE \;\;\;\; request[msg.sender][\_to] = \textbf{true};
\STATE \}
\end{algorithmic} 
\small
\end{algorithm}

The function for the second stage of the framework which is \textit{giving consent} is given in Algorithm 4. The function needs the sender address as an only input from which the receiver will receive tokens since there can be multiple consent requests from different senders. The proposed smart contract uses a global mapping variable named as \textit{consent} to manage the consents between senders and receivers. The possible keys and the values of the mapping are the same with the previous mapping. The value of true means that there is a consent from the receiver to the sender and vice versa. Once the function is called, the corresponding value in the mapping is set to true.

\begin{algorithm} 
\small
\caption{Receiver Giving Consent}  
\begin{algorithmic}[1]
\STATE \textbf{function} giveConsent(\textbf{address} \_from) \textbf{public} \{
\STATE \;\;\;\; \textbf{require}(request[\_from][msg.sender] == \textbf{true});
\STATE \;\;\;\; consent[\_from][msg.sender] = \textbf{true};
\STATE \}
\end{algorithmic} 
\small
\end{algorithm}

The function for the third stage of the framework as \textit{private depositing}, is given in Algorithm 5, which is more complex than the functions of the first two stages. It takes five different inputs as (1) the receiver address to send tokens, (2) the hash for transaction amount, (3) the hash for the next balance of the sender, (4) the encrypted message between the sender and the receiver, and finally (5) the zero-knowledge proof. There are some important points that should be immediately noted. The first point is that the current balance of the sender (i.e. \textit{balanceHash[msg.sender]}) is not provided by the sender itself, rather it is already stored in the private token smart contract. Otherwise, the sender would be able to generate a zero-knowledge proof for whatever balance the sender wishes, which would result in cheating. The second point is that the encrypted message constitutes private channel so that the sender can convey secret information including transaction amount and corresponding blind factor directly to the receiver. It is encrypted by the public key of the receiver. 

Inside the function, the consent from the receiver to the sender is checked at first and reverted if there is no consent. Later, the function of \textit{verifyTx} is called with the proof and other necessary inputs from an external smart contract that is already generated by the Zokrates tool and deployed to the blockchain. The main responsibility of the external function is to validate the zero-knowledge proof with the verifying key stored in the external contract. If the sender proof is successfully passed, the balance of the sender is updated accordingly. Moreover, our private token smart contract uses a global mapping named as \textit{allowance} so that the receiver later can collect the exact amount of tokens the sender sends. Otherwise, the receiver would be able to collect whatever amount of tokens the receiver wishes.  

\begin{algorithm*} 
\small
\caption{Private Depositing}  
\begin{algorithmic}[1]
\STATE \textbf{function} privateDeposit(\textbf{address} \_to, \textbf{uint256} amountHash, \textbf{uint256} nextBalanceHash, \textbf{string} encryptedMessage, \textbf{Proof} proof) \textbf{public} \{
\STATE \;\;\;\; \textbf{require}(consent[msg.sender][\_to] == \textbf{true});
\STATE \;\;\;\; isProofCorrect = verifier.verifyTx(proof, amountHash, balanceHash[msg.sender], nextBalanceHash);
\STATE \;\;\;\; \textbf{if}(isProofCorrect) \{ 
\STATE \;\;\;\;\;\;\;\; balanceHash[msg.sender] = nextBalanceHash;
\STATE \;\;\;\;\;\;\;\; allowance[msg.sender][\_to] = amountHash;
\STATE \;\;\;\;\;\;\;\; encryptedMessages[msg.sender][\_to] = encryptedMessage;
\STATE \;\;\;\; \} 
\STATE \} 
\end{algorithmic} 
\small
\end{algorithm*}

The function for the fourth and final stage of the framework as \textit{private withdrawing}, is given in Algorithm 6. It takes four different inputs as (1) the sender address to receive tokens, (2) the hash for transaction amount, (3) the hash for the next balance of the receiver, and (4) the zero-knowledge proof. Similar to the same concern in the previous stage, the current balance of the receiver is not provided by the receiver itself, rather it is fetched from the contract. Inside the function, the consent and the allowance between the sender and the receiver is checked. Later, the zero-knowledge proof is validated using the the function of \textit{verifyTx} from the external smart contract already generated and deployed to the blockchain. If the receiver proof is successfully passed, the balance of the receiver is updated accordingly. The allowance is also set to zero so that the receiver would not be able to unfairly replay and collect the tokens again. 

\begin{algorithm*} 
\small
\caption{Private Withdrawing}  
\begin{algorithmic}[1]
\STATE \textbf{function} privateWithdraw(\textbf{address} \_from, \textbf{uint256} amountHash, \textbf{uint256} nextBalanceHash, \textbf{Proof} proof) \textbf{public} \{
\STATE \;\;\;\; \textbf{require}(consent[\_from][msg.sender] == \textbf{true});
\STATE \;\;\;\; \textbf{require}(allowance[\_from][msg.sender] == amountHash);
\STATE \;\;\;\; isProofCorrect = verifier.verifyTx(proof, amountHash, balanceHash[msg.sender], nextBalanceHash);
\STATE \;\;\;\; \textbf{if}(isProofCorrect) \{ 
\STATE \;\;\;\;\;\;\;\; balanceHash[msg.sender] = nextBalanceHash;
\STATE \;\;\;\;\;\;\;\; allowance[msg.sender][\_to] = 0;
\STATE \;\;\;\; \}
\STATE \} 
\end{algorithmic} 
\small
\end{algorithm*}

\subsection{Web User Interface Design}

The direct interaction with our private token smart contract involving with can be difficult for the end users due to zero-knowledge proof generation/verification, asymmetric encryption and other complex operations. Hence, we have designed a web user interface where the end users can use it through their browsers without any trusted setup and configuration. The only condition required for the application to run is the Metamask extension \cite{metamask} bridging the interface to the private token smart contract. Overall, our private token transfer system (PTTS) is autonomous decentralized exchange application where all data is stored in the blockchain without any centralized server. Any end users who download the application from the GitHub page can run and test it on their browsers. 

Our web user interface supports several functionalities including (1) ordinary private token deployment in the first stage, (2) getting consent in the second stage, (3) giving consent in the third stage, (4) depositing tokens in the fourth stage and (5) withdrawing tokens in the fifth and the last stage of the framework proposed. In the first functionality, the users are simply prompted to provide appropriate token name and symbol as strings. Once the button is clicked, the interface communicates with Ethereum through Metamask, deploys a new private token and returns that token address and transaction hash to the user. The token owner can share the token address so that the users can involve with private transfer over that token. The user interface is given in Fig 2. 

\begin{figure}[htbp]
\centerline{\fbox{\includegraphics[width=0.7\textwidth]{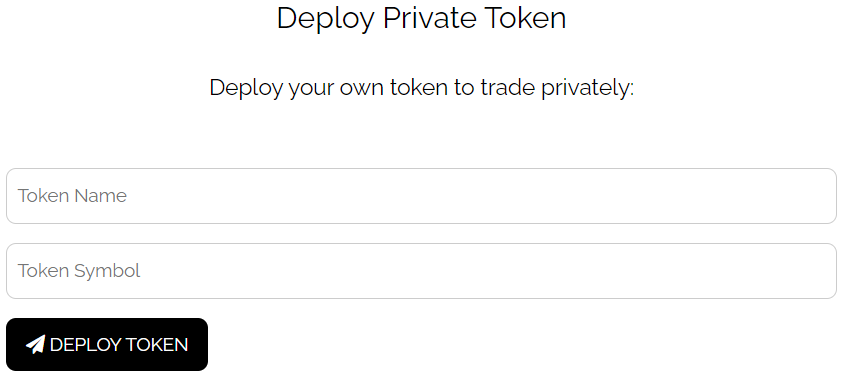}}}
\caption{The User Interface for Private Token Deployment}
\end{figure}

The second functionality is getting consent where the users provide the token address they want to interact and the receiver address to get consent from. The corresponding user interface is given in Fig 3. Similarly, the third functionality is giving consent where the users provide the token address and the sender address to give consent to. The public key of the receiver is restored from the resulting transaction at that stage as well once the button is clicked. The public key will be later used by the sender in order to encrypt the transaction amount and the corresponding blind factor during the next stage. The corresponding user interface is given in Fig 4. 

\begin{figure}[htbp]
\centerline{\fbox{\includegraphics[width=0.7\textwidth]{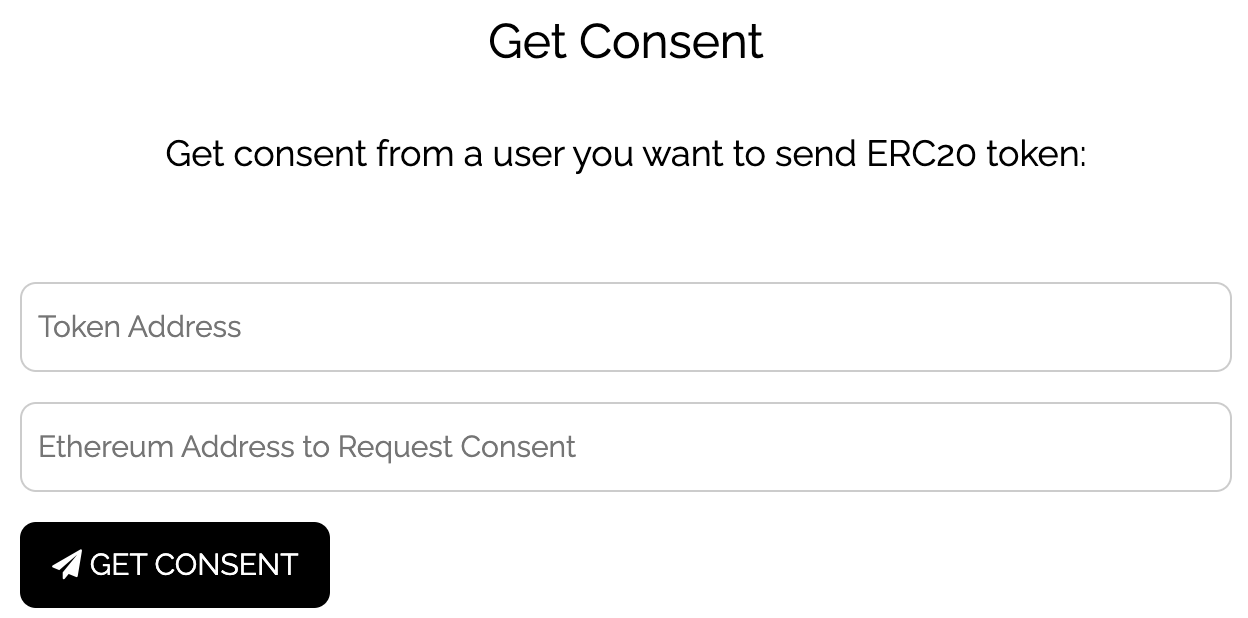}}}
\caption{The User Interface for Getting Consent}
\end{figure}

\begin{figure}[htbp]
\centerline{\fbox{\includegraphics[width=0.7\textwidth]{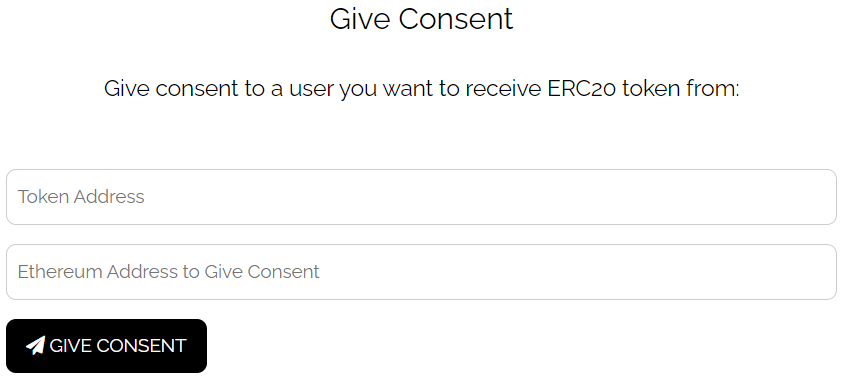}}}
\caption{The User Interface for Giving Consent}
\end{figure}

The fourth functionality is depositing private tokens which is more complex with respect the type and number of operations performed in the background. It prompts the sender to enter five inputs as (1) token address, (2) the receiver address to send tokens, (3) the transaction amount which will be encrypted, (4) current balance of the sender and (5) blind factor for the current balance. Once the button is clicked, the blind factors required for the transaction amount and the next sender balance are generated while the third blind factor is already provided by the sender. Later, the message from the sender to the receiver is encrypted using the receiver public key including the transaction amount and the corresponding blind factor. The hashes for the transaction amount, the current sender balance and the next sender balance to be used during Zokrates proof generation are computed. When ready, Zokrates generates the proof in the browser. If the proof is validated in our smart contract, the tokens are sent and the sender balance is updated. The user interface is given in Fig 5.

\begin{figure}[htbp]
\centerline{\fbox{\includegraphics[width=0.7\textwidth]{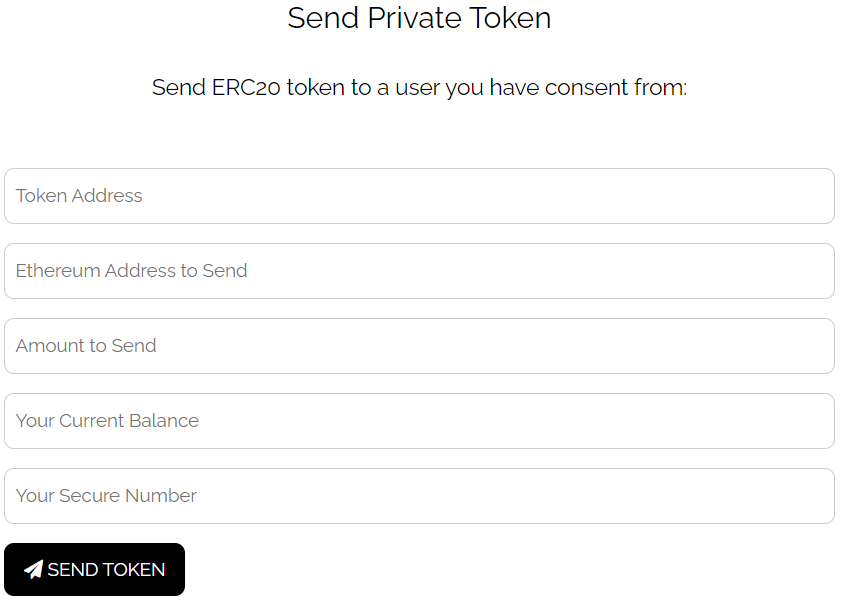}}}
\caption{The User Interface for Private Depositing}
\end{figure}

The fifth functionality is withdrawing private tokens where it prompts the receiver to enter five inputs as (1) token address, (2) the sender address to receive tokens, (3) the private key which will be used only inside the browser for decryption, (4) current balance of the receiver and (5) blind factor for the current balance. Once the button is clicked, only a  blind factor for the next receiver balance is generated since the blind factor for the current receiver balance is provided by the receiver while the the blind factor for the transaction amount is decrypted from the encrypted message of the sender. After necessary hashes are computed, Zokrates generates the proof in the browser. If it successfully passes in our smart contract, the tokens are received and the receiver balance is updated. The user interface is given in Fig 6. The complete architecture including the zero-knowledge proofs, the smart contract and the web user interface is also given in Fig 7.

\begin{figure}[htbp]
\centerline{\fbox{\includegraphics[width=0.7\textwidth]{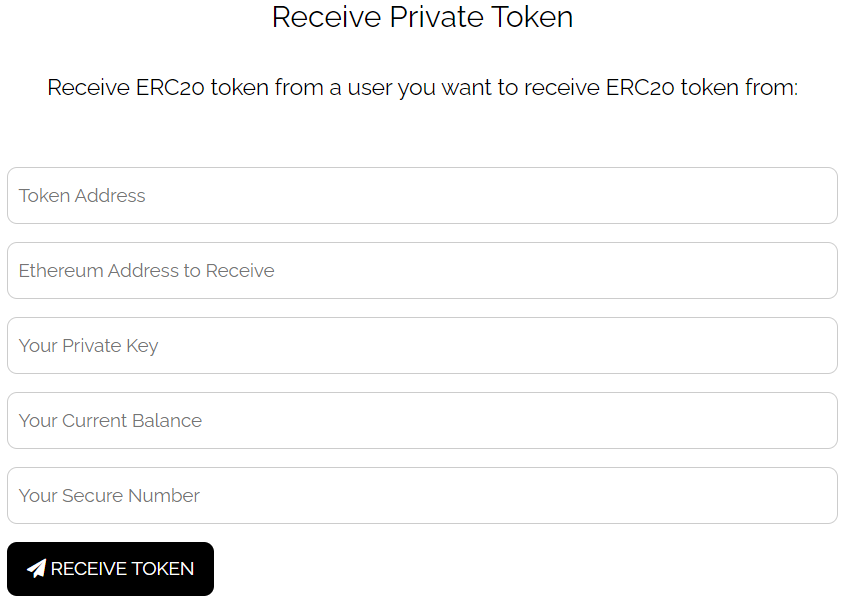}}}
\caption{The User Interface for Private Withdrawing}
\end{figure}

\begin{figure*}[htbp]
\centerline{\includegraphics[width=1.0\textwidth]{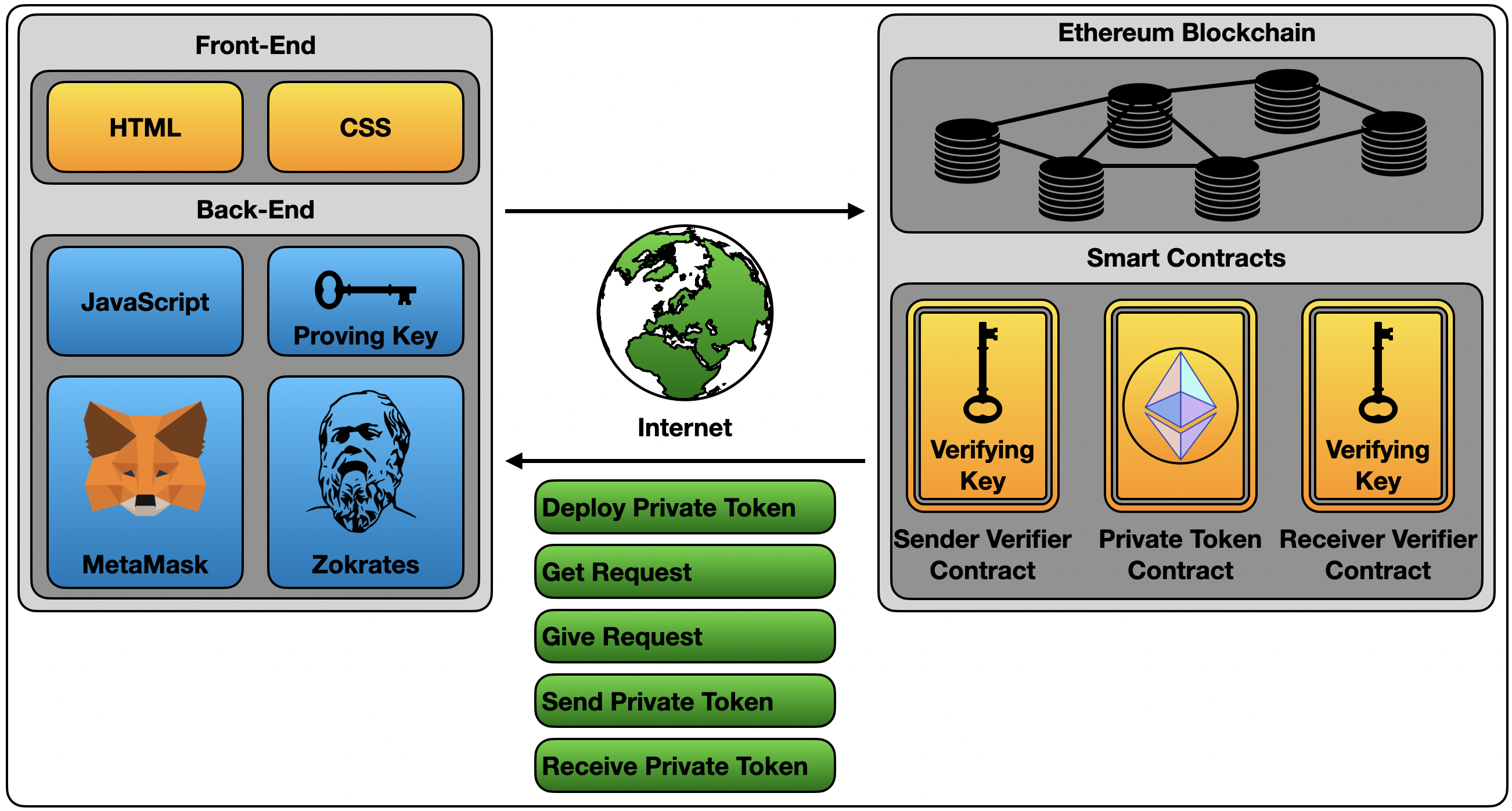}}
\caption{The Private Token Transfer System Architecture}
\end{figure*}

\section{Security Analysis of Private Token Transfer System}

\subsection{Replay Attack}

The replay attack refers to duplication of a valid transaction in the blockchain and maliciously replaying it again. A successful replay attack may eventually result in the loss of funds and assets. Our private token transfer system is robust against the replay attack. When the stages are individually analysed, it is easy to see that the \textit{getting consent} and the \textit{giving consent} stages of the framework are handshake stages between sender and receiver where certain values in the  mappings are set to true in the contract. Even if the transactions corresponding to these stages are replicated again, the same values in the mappings are set to true, which will be safe. However, the transactions for the \textit{private depositing} and the \textit{private withdrawing} stages need more elaboration. 
 
For a moment, suppose that our system did not use any blind factors for the transaction amount, the current balance and the next balance. Then, the zero-knowledge proof in the third and the fourth stages would be generated specifically for three values as tuple $<$$tx\_amnt, curr\_bal, next\_bal$$>$, respectively. The generated proof correctly would not be valid for any other tuple. However, it would be relatively easy to find the same tuple again inside the blockchain where the proof would be able to be maliciously validated. In our framework, the zero-knowledge proof is generated specifically for three values along with their blind factors as tuple $<$$tx\_amnt, tx\_amnt\_bf, curr\_bal, curr\_bal\_bf, next\_bal, next\_bal\_bf$$>$. The blind factors are randomly generated in the browser using 52-bits of randomness where the collisions for each blind factor should be around $2^{26} \approx 64$ million with respect to the popular birthday paradox \cite{birthday}. When considered for three blind factors combined, it is practically infeasible to find the same tuple again in the blockchain. Therefore, the proof generated for a specific tuple will not be valid any more once it is validated in the transaction. 

\subsection{Balance Range Privacy Attack}

\subsubsection{Problem Definition}

The assumption that cryptographically secure private token transfer systems is completely robust to the external privacy attacks might not be always correct. In certain situations where some of the private transactions are deliberately leaked out or sold to the specific actors of the blockchain including exchange services and governments, it may be possible to extract meaningful information about user balances. Such organizations may progressively collect the private transactions performed in the blockchain by offering certain benefits to the users. From the perspective of rational users, it may be preferable to leak/sell the transactions if the benefits outweigh the private transactions themselves. 

In this problem, we have asked the following question: "Is it practically possible to attack the system in order to estimate the feasible balance ranges if some of the private transactions are already known by an actor?". By looking at various tests that we have performed, we have also asked: "What is the relation between the balance range estimation certainty and the ratio of private transactions collected with respect to the total transactions performed in the blockchain?". It is obvious that this certainty would be 100\% if all the private transactions were collected. In our experimental study, we have performed tests with increasing collection ratio in order to understand that relation. In case the experiments are promising, it may be better to examine the security of today's private token transfer systems again from this standpoint.

\subsubsection{Minimum Cost Flow Network Design}

The balance range privacy attack problem can be solved by two executions of minimum cost flow network algorithm with certain cost assignments on the edges. Let $G = (V, E)$ be a directed graph where $V$ is the set of nodes (i.e.  users) and $E$ is the set of edges (i.e. transactions). $E^*$ is the subset of transactions that are already leaked/sold to a particular organization in the blockchain. Each node $i$ is associated with a supply value $b_i$ where the source node $S$ supplies all the tokens while the sink node $T$ demands these tokens and the rest of 
the nodes have $b_i$ set to 0. The node corresponding to the user whose balance is in our focus is denoted as $U$. Each edge between the nodes $i$ and $j$ is also associated with lower/upper bounds $[l_{ij}, u_{ij}]$  on the feasible flow that can pass through the edge
and a cost value $c_{ij}$ to be used in the objective function. For each edge corresponding to a leaked transaction, the lower and upper bounds are the same and equal to the transaction amount $v_{ij}$. For the remaining non-leaked transactions, we take $l_{ij}$ as 0 and $u_{ij}$ as the total token supply. The decision variable (i.e. flow) on the edge is denoted as $x_{ij}$. 

In the first minimum cost flow network to be solved, the costs of all the edges, except the edge between the user in focus and the sink node, are set to zero. However, the edge between the user in focus and the sink node is set to $-1$ which means that the network solver will try to flow as many tokens as possible on this edge in order to minimize the overall cost of the network. Similarly, in the second minimum cost flow network to be solved, the edge between the user in focus and the sink node is set to $+1$ while all the other edges are still zero. In this case, the network flow solver will try to flow  as few tokens as possible through 
this edge in order not to increase the overall cost of the network. In the formulation of the two network flow problems, the two objectives are:

\begin{equation}
min. \sum_{(i, j) \in E} c_{ij} x_{ij} = -c_{UT} x_{UT}
\end{equation}

\begin{equation}
min. \sum_{(i, j) \in E} c_{ij} x_{ij} = +c_{UT} x_{UT}
\end{equation}

\noindent
which are both subject to the following same constraints:

\begin{align}
s.t. & \sum_{(i, j) \in E} x_{ij} - \sum_{(j, i) \in E} x_{ij} = b_i, \forall i \in V \nonumber \\
& l_{ij} \leq x_{ij} \leq u_{ij} , \forall (i, j) \in E \nonumber \\
& l_{ij} = u_{ij} = v_{ij}, \forall (i, j) \in E^* \nonumber \\
& b_{S} = b_{total} \nonumber \\
& b_{T} = -b_{total} \nonumber \\
\end{align}

\noindent
Here, $b_{total}$ is the total token supply to the blockchain. The eqns. (1) and (2) are the objectives of the first and the second minimum cost flow network problems respectively while the eqn. (3) are the common constraints. The illustrative graphs for the networks are given in Fig 8. We assume that for the balance of the user in focus, there is only one feasible region that is limited by minimum and the maximum feasible values found. The correctness of this assumption will be shown in the next subsection.

\begin{figure}
\centering
\begin{subfigure}[b]{0.6\textwidth}
   \includegraphics[width=1\linewidth]{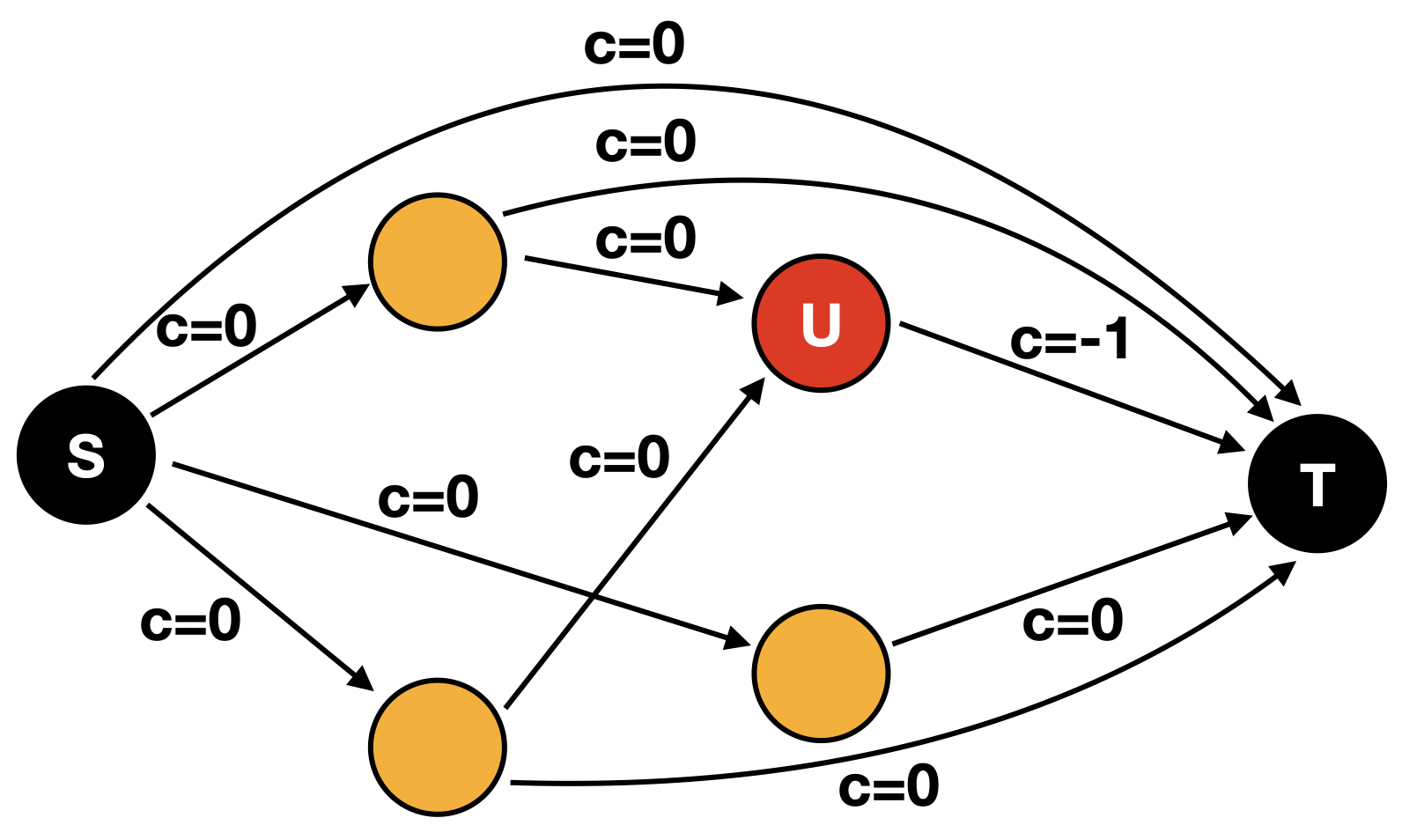}
   \caption{}
\end{subfigure}

\begin{subfigure}[b]{0.6\textwidth}
   \includegraphics[width=1\linewidth]{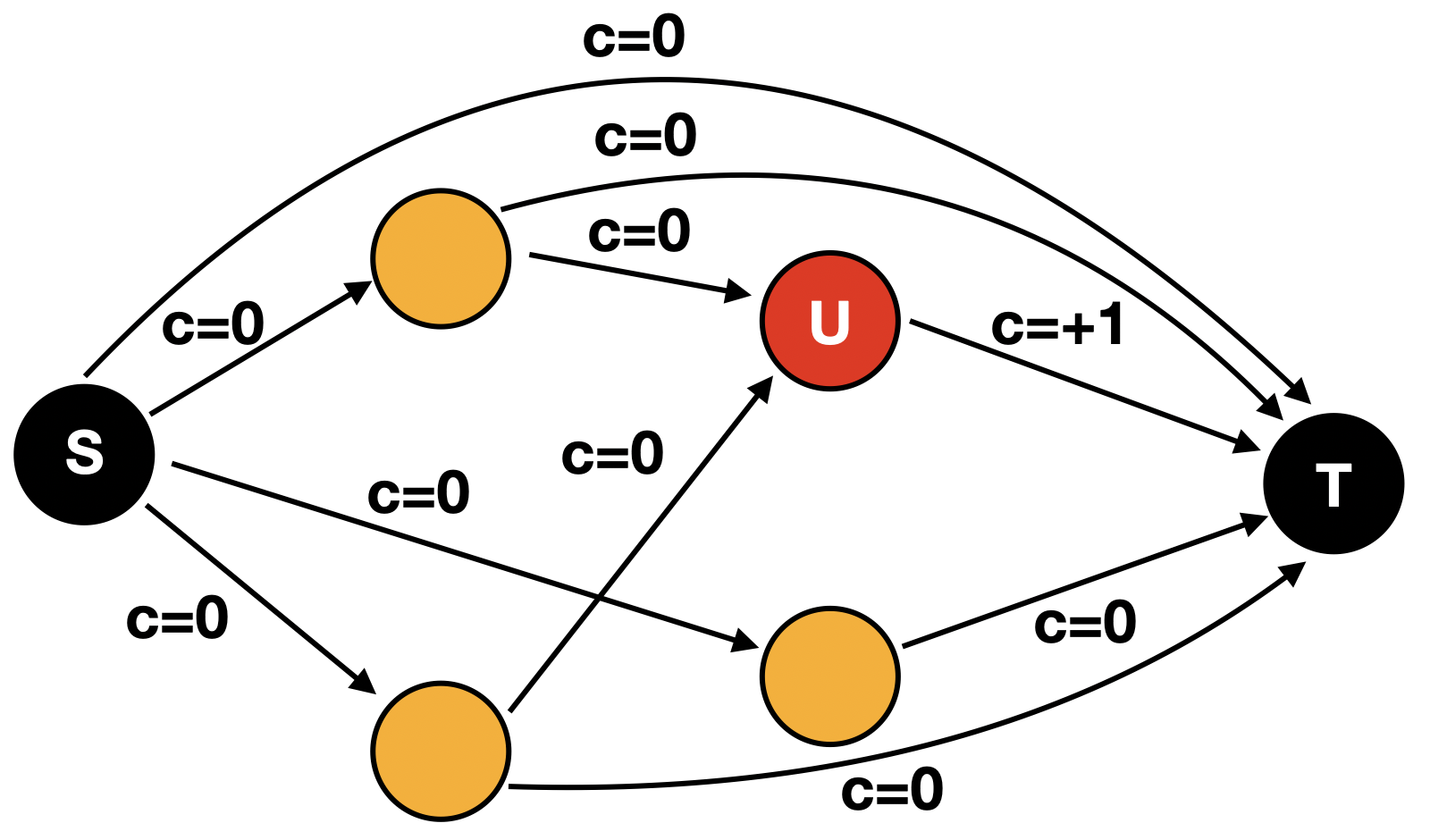}
   \caption{}
\end{subfigure}

\caption[]{Balance Range Disclosure Networks: (a) Minimum cost flow network for maximum feasible balance, (b) Minimum cost flow network for minimum feasible balance. S is the source node, T is the sink node.}
\end{figure}

\subsubsection{Contiguity in Balance Range}

By solving two minimum cost flow networks with +1 and -1 cost  assignments on certain edges, we find the minimum and the maximum possible balances that a specific user may have. However, such an approach only becomes valid as long as all the balances between the minimum value and the maximum value are contiguously feasible. Otherwise, we would end up with multi-piece feasible regions in-between rather than single one-piece feasible region. Based upon this balance contiguity issue, the goal here is to show whether all the balances rested between the minimum and the maximum feasible balances are also feasible to validate the correctness of our approach. 

For this issue, it should be first noted that the minimum cost flow network model of our balance range privacy attack problem is a linear optimization problem where the constraints are in the form of $A x \leq b$. Any linear optimization problem is also a convex problem which results in a convex feasible region. According to the definition of convexity, any convex combination $\forall \alpha \in [0, 1]$ of any two arbitrary points $x_1, x_2 \in X$ taken from the feasible region $X$ must reside in the same feasible region \cite{boyd2004}: 
\begin{align}
Ax_3 = \alpha Ax_1 + (1-\alpha)Ax_2 &\leq \alpha b + (1-\alpha)b = b \nonumber \\
Ax_3 &\leq b
\end{align}

\noindent
where $x_3 = \alpha x_1 + (1-\alpha)x_2$. Based on these facts, it can be said that any balance between the minimum and the maximum feasible balances is also feasible.

\section{Experimental Evaluation}

\subsection{Experimental Setup}

The private token transfer system consists of smart contracts in the blockchain-side and web application in the client-side. The frontend design of the web application is based on HTML and CSS languages while the backend design of the web application is based on Javascript language. For the backend design, $(i)$ the \textit{ethers} library \cite{ethers} is used in order to bridge the web application to Ethereum, $(ii)$ the \textit{eccrypto} library \cite{eccrpyto} is used to encrypt and decrypt with elliptic curve cryptography, $(iii)$ the Zokrates privacy-preserving tool is used \cite{zokrates} to generate verifiable proofs and finally $(iv)$ \textit{browserify} \cite{browserify} and \textit{webpack} \cite{webpack} bundlers are used to be able to run functions in the browser. For the blockchain-side, the smart contracts in Ethereum are implemented using the Solidity language \cite{solidity} and compiled with 0.8.0-version of Remix compiler \cite{remix}. 

For the balance range privacy attack, the users and the transactions between these users are randomly generated with respect to the number of users and the number of transactions parameters, respectively. The transactions to be disclosed to the adverserial entities are also randomly selected among all the transactions. The values for the network solution times and the goodness rate metrics are taken as the average of 20 different runs and in each run, balance range of a randomly selected user is attacked (i.e. investigated). During the experiments, $(i)$ the number of addresses changes in the range of 100 and 1,000,000, $(ii)$ the number of transactions changes in the range of 100 and 1,000,000 and $(iii)$ the default total token supply is 1,000,000 and default transaction leakage ratio is 0.5. We have employed Google OR-Tools \cite{or} and the \textit{Parallel Network Simplex} algorithm \cite{pns} to solve the minimum cost flow network. All the tests are performed upon MacBook Pro Notebook with a 2.6 GHz Intel Core i7 processor, 16 GB memory and 6 cores. The implementations for the private token transfer system and the balance range disclosure attack are available in our GitHub page \cite{github}.

\subsection{Experiments For Private Token Transfer System}

It has been stated in the problem definition of our private token transfer system that four different requirements are required including privacy, public verifiability, authentication and non-interactivity. When these requirements are evaluated under the findings of the proposed architecture, it can be said that our system satisfies the first requirement since no user balances and transaction amounts are revealed while depositing or withdrawing the tokens from the smart contract. Meanwhile, it satisfies the second requirement since the validator nodes in the network can still validate the correctness of the private transactions through zero-knowledge proof protocols. It also satisfies the third requirement since two-way handshake must be performed between the sender and the receiver before the actual token transfer and therefore, no unauthorized receiver can withdraw the tokens from the smart contract. Furthermore, the receiver, even it is authorized, cannot withdraw an amount of token more than deposited. Lastly, our system partially satisfies the fourth requirement because of the two-way handshake. In the future work, we are planning to work on this issue to achieve full non-interactivity.

\begin{table*}[htbp]
\caption{The gas costs of the functions in our private token smart contract (Gas Price: 1.5 Gwei, 1 Ether: \$ 1,197.66 )}
\footnotesize
\begin{center}
\begin{tabular}{|l|r|c|c|}
\hline
\cellcolor{gray!50} \textbf{Function} & \cellcolor{gray!50} \textbf{Gas Units} &  \cellcolor{gray!50} \textbf{Gas Cost (Ether)} & \cellcolor{gray!50} \textbf{Gas Cost (USD)} \\   
\hline
\textbf{Deployment} & 4,557,726 & 0.00683658 & \$ 8.19  \\
\hline
\textbf{Get Consent} & 44,300 & 0.00006645 & \$ 0.08  \\
\hline
\textbf{Give Consent} & 183,344 & 0.00027501 & \$ 0.33  \\
\hline
\textbf{Deposit Tokens} & 2,060,133 & 0.00309019 & \$ 3.70  \\
\hline
\textbf{Withdraw Tokens} & 1,699,399 & 0.00253816 & \$ 3.04  \\
\hline
\end{tabular}
\end{center}
\footnotesize
\end{table*}

The gas cost consumptions of the functions in our private token smart contract corresponding to the four available stages in our system are presented in Table 1. These functions include private token deployment, getting consent from the receiver, giving consent to the sender, depositing private tokens with zero-knowledge proof and withdrawing private tokens with zero-knowledge proof. As of 27/06/2022, the gas price is approximately 1.5 Gwei and the exchange rate from ether to dollar is \$ 1,197.66 . As seen in Table 1, the most expensive function among all the functions is contract deployment with over 4,5 million gas units corresponding to over \$ 8. This amount seems feasible and reasonable since the smart contract has two external function calls to validate the zero-knowledge proofs submitted from the web user interface including one call for sender proofs and another call for receiver proofs. 

Among the other functions given in Table 1, depositing and withdrawing tokens consume the higher gases due to on-chain zero-knowledge proof validations. Overall, the sender has to pay approximately \$ 3.78  (= \$ 0.08  + \$ 3.70 ) to deposit tokens to the smart contract while the receiver has to pay approximately \$ 3.37  (= \$ 0.33  + \$ 3.04 ) to withdraw those tokens. Another important point to consider is that although the proof validation process is fast, the proof generation process of Zokrates is the principal bottleneck in our system. It even slows down more while using the web user interface in the browser with respect to the command-line interface. The proof generation times on average take up to nearly 147 seconds for the senders and 145 seconds for the receivers. It is natural to observe such a slight increase in the sender proof generation times because as discussed previously in Algorithm 1 and 2, it checks four conditions during that process while the receiver proof generation checks only three conditions.

\subsection{Experiments For Balance Range Disclosure Attack}
\label{experimentsbalanceattack}

\subsubsection{For Number of Addresses and Transactions}

\begin{table*}[htbp]
\caption{The network solution times and goodness rates for different number of addresses and transactions. Transaction leakage ratio is taken as 0.5 .}
\scriptsize
\begin{center}
\begin{tabular}{|l|l|r|r|r|r|r|}
\hline
\multicolumn{2}{|l|}{\cellcolor{black!100}} & \multicolumn{5}{c|}{\cellcolor{gray!50} \textbf{Number of Transactions}} \\   
\hline
\multicolumn{1}{|c|}{\cellcolor{gray!50} \textbf{Number of}} & \multicolumn{1}{c|}{ \cellcolor{gray!50} \textbf{Metrics}} &  \textbf{100} &  \textbf{1,000} &  \textbf{10,000} &  \textbf{100,000} &  \textbf{1,000,000} \\   

\multicolumn{1}{|l|}{\cellcolor{gray!50} \textbf{Addresses}} & \multicolumn{1}{c|}{ \cellcolor{gray!50}} &  &  &  &  & \\ 

\hline
\textbf{100} & \textbf{Time - Google OR (sec)} & $<$0.01 & $<$0.01 & 0.01 & 0.10 & 1.05 \\ \textbf{100} & \textbf{Time - PNS (sec)} & $<$0.01 & $<$0.01 & 0.02 & 0.19 & 1.89 \\ \textbf{100} &  \textbf{Goodness Rate} & 0.89 & 0.06 & 0.00 & 0.00 & 0.00 \\ 
\hline
\textbf{1,000} & \textbf{Time - Google OR (sec)} & $<$0.01 & $<$0.01 & 0.02 & 0.13 & 1.41 \\ \textbf{1,000} & \textbf{Time - PNS (sec)} & $<$0.01 & $<$0.01 & 0.03 & 0.21 & 2.03 \\ \textbf{1,000} & \textbf{Goodness Rate} & 0.94 & 0.90 & 0.01 & 0.00 & 0.00 \\
\hline
\textbf{10,000} & \textbf{Time - Google OR (sec)} & 0.02 & 0.02 & 0.10 & 0.35 & 2.37 \\ \textbf{10,000} & \textbf{Time - PNS (sec)} & 0.05 & 0.05 & 0.08 & 0.42 & 2.54 \\ \textbf{10,000} & \textbf{Goodness Rate} & 0.98 & 0.98 & 0.91 & 0.03 & 0.00 \\
\hline
\textbf{100,000} & \textbf{Time - Google OR (sec)} & 0.15 & 0.16 & 0.23 & 2.97 & 9.83 \\ \textbf{100,000} & \textbf{Time - PNS (sec)} & 0.49 & 0.55 & 0.62 & 1.45 & 24.19 \\ \textbf{100,000} & \textbf{Goodness Rate} & $>$0.99 & 0.98 & 0.98 & 0.89 & 0.02 \\
\hline
\textbf{1,000,000} & \textbf{Time - Google OR (sec)} & 1.61 & 1.72 & 2.14 & 4.20 & 185.72 \\ \textbf{1,000,000} & \textbf{Time - PNS (sec)} & 7.75 & 8.14 & 10.11 & 11.96 & 425.82 \\ \textbf{1,000,000} & \textbf{Goodness Rate} & $>$0.99 & $>$0.99 & $>$0.99 & $>$0.99 & 0.98 \\
\hline
\end{tabular}
\end{center}
\scriptsize
\end{table*}

We have evaluated the performance of our approach to perform the balance range disclosure attack with respect to two different metrics as the elapsed time to solve the minimum cost flow networks and the balance goodness rate. Although the calculation of the first metric is simple where the start time of the network solver is subtracted from its finish time, the second metric needs more elaboration. The second metric specifically designed for our approach measures how much the balance range for the user in focus is narrowed down by considering the maximum possible range from zero to the token supply. The formula that we used for the second metric, which we call $goodness$ is as follows:

\begin{equation}
goodness = 1 - \frac{(max\_value - min\_value)}{token\_supply}
\end{equation}

\noindent
where $min\_value$ and $max\_value$ are the values found by our approach while the $token\_supply$ is the maximum available tokens supplied to the blockchain users. This formulation results in goodness rate between zero and one. In the best case where the minimum value found, the maximum value found and the actual balance are all equal to each other, the goodness rate becomes one. On the other hand, in the worst case where the minimum value found is zero since it cannot be negative and the maximum value found is equal to the token supply, then the goodness rate becomes zero. It should also be noted that there is no case in our experiments where the actual balance of the user is found to be out of range computed by our approach 
which is as expected. 

\begin{figure}[htbp]
\centerline{\fbox{\includegraphics[width=1\textwidth]{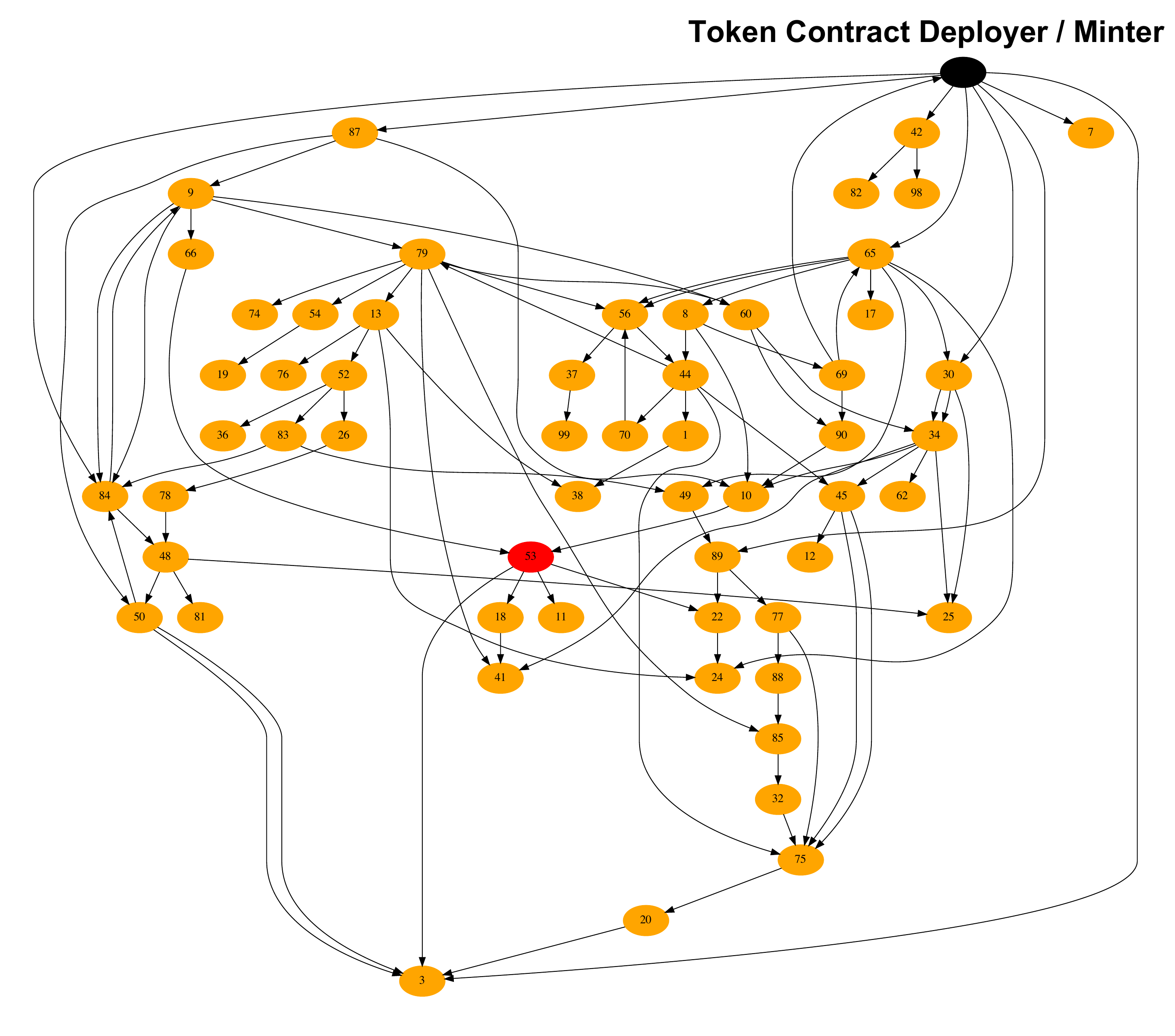}}}
\caption{The private transaction graph: (black): source and sink addresses, (red): address in focus, (orange): ordinary addresses involving with private transactions}
\end{figure}

\begin{figure}[htbp]
\centerline{\fbox{\includegraphics[width=1\textwidth]{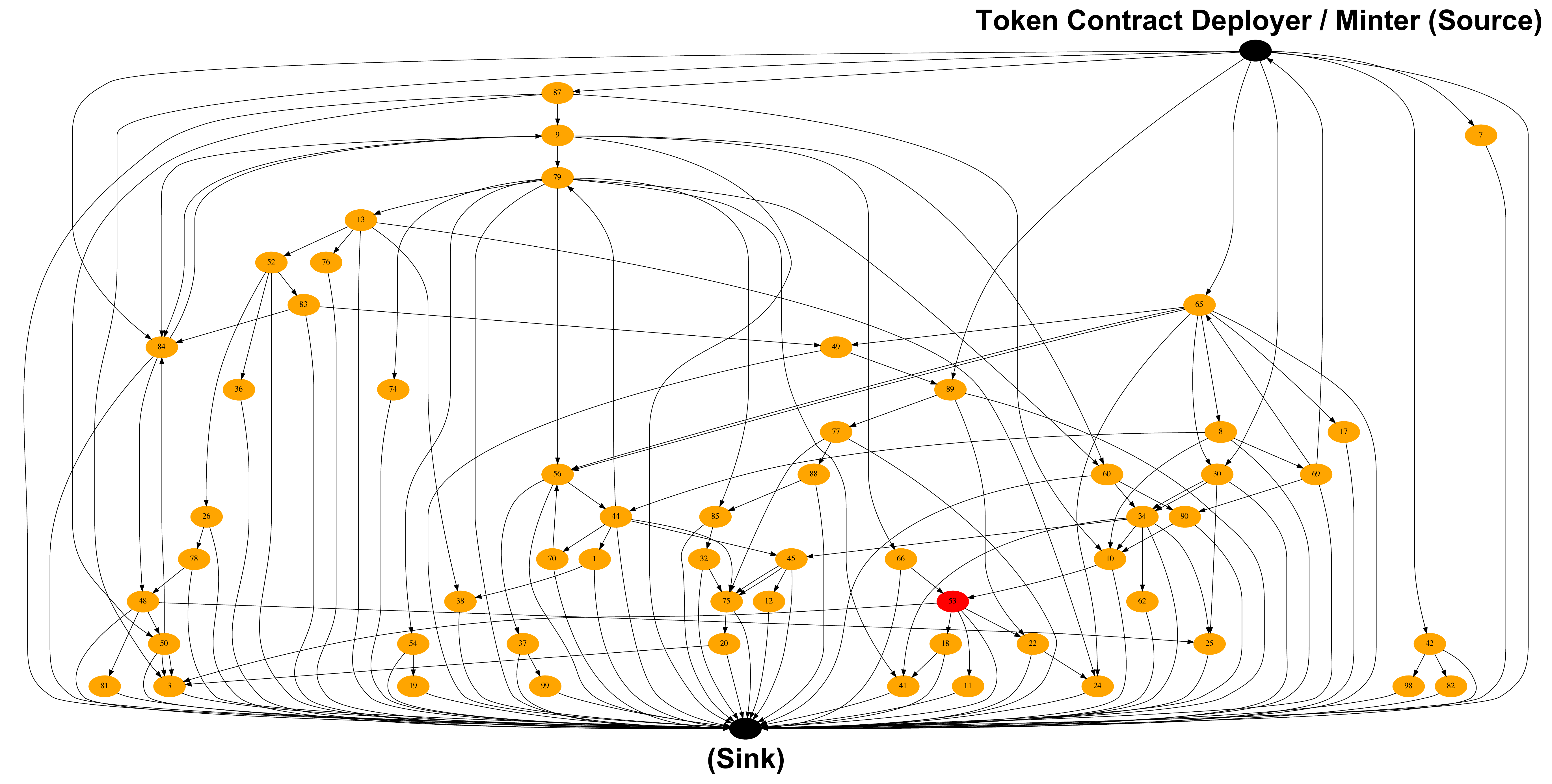}}}
\caption{The minimum cost flow network: (black): source and sink addresses, (red): address in focus, (orange): ordinary addresses involving with private transactions}
\end{figure}

An example transaction graph and the resulting minimum cost flow network from our experiments are shown in Fig. 9 and 10 where the black nodes are source and sink addresses; the red nodes are the address in focus and the orange nodes are the ordinary addresses. The node and edge labels are omitted for simplicity. The experimental results for the number of addresses and the number of transactions are given in Table 2 where total token supply is assumed to be  1,000,000 and the transaction leakage ratio is assumed to be 0.5 which means that half of all transactions are leaked to the malicious organization. According to the results, there can be several deductions to be made. The first metric says that the amount of time to solve the networks increases with increasing number of addresses and transactions, as expected. However, in most of the cases, the networks can be solved under two seconds. The most extreme case is for 1,000,000 addresses and 1,000,000 transactions, which requires more than two minutes to be solved. 

The second metric brings up the concept of transaction to address ratio which shows how many transactions the users are involved with on average. In the scenarios where transaction to address ratio is higher, it harder to trace the transactions and estimate the balance ranges correctly since only half of the transactions are leaked out. We have seen these effects clearly for 100 addresses and 10,000 transactions (transaction to address ratio is 100) where the goodness rate is zero. It means that it is possible for the user balances to vary between zero and the total token supply. On the other hand, in scenarios where the users are involved with fewer transactions as shown Table 2 (i.e. transaction to address ratio is less than one), it may be easier to have better estimation of 
the balance ranges. 

\subsubsection{For Transaction Leakage Ratio}

\begin{table*}[htbp]
\caption{The network solution times and goodness rates for different transaction leakage ratios}
\scriptsize
\begin{center}
\resizebox{\columnwidth}{!}{%
\begin{tabular}{|l|l|r|r|r|r|r|}
\hline
\multicolumn{2}{|l|}{\cellcolor{black!100}} & \multicolumn{5}{c|}{\cellcolor{gray!50} \textbf{Number of Addresses/}} \\
\multicolumn{2}{|l|}{\cellcolor{black!100}} & \multicolumn{5}{c|}{\cellcolor{gray!50} \textbf{Number of Transactions}} \\

\hline
\multicolumn{1}{|c|}{\cellcolor{gray!50} \textbf{Transaction}} & \multicolumn{1}{c|}{ \cellcolor{gray!50} \textbf{Metrics}} & \textbf{100/} & \textbf{1,000/} & \textbf{10,000/} & \textbf{100,000/} & \textbf{1,000,000/} \\   
\multicolumn{1}{|l|}{\cellcolor{gray!50} \textbf{Leakage Ratio}} & \multicolumn{1}{c|}{\cellcolor{gray!50}} & \textbf{100} & \textbf{1,000} & \textbf{10,000} & \textbf{100,000} & \textbf{1,000,000} \\   
\hline
\textbf{0.2} & \textbf{Time - Google OR (sec)} & $<$0.01 & $<$0.01 & 0.13 & 2.88 & 135.20 \\ \textbf{0.2} & \textbf{Time - PNS (sec)} & $<$0.01 & $<$0.01 & 0.08 & 1.19 & 464.14 \\ \textbf{0.2} & \textbf{Goodness Rate} & 0.38 & 0.36 & 0.34 & 0.30 & 0.29 \\ 
\hline
\textbf{0.4} & \textbf{Time - Google OR (sec)} & $<$0.01 & $<$0.01 & 0.15 & 3.55 & 142.29 \\ \textbf{0.4} & \textbf{Time - PNS (sec)} & $<$0.01 & $<$0.01 & 0.08 & 2.17 & 574.74 \\ \textbf{0.4} & \textbf{Goodness Rate} & 0.84 & 0.69 & 0.76 & 0.63 & 0.82 \\
\hline
\textbf{0.6} & \textbf{Time - Google OR (sec)} & $<$0.01 & $<$0.01 & 0.07 & 1.14 & 26.93 \\ \textbf{0.6} & \textbf{Time - PNS (sec)} & $<$0.01 & $<$0.01 & 0.08 & 1.55 & 293.54 \\ \textbf{0.6} & \textbf{Goodness Rate} & 0.86 & 0.95 & $>$0.99 & $>$0.99 & $>$0.99 \\
\hline
\textbf{0.8} & \textbf{Time - Google OR (sec)} & $<$0.01 & $<$0.01 & 0.05 & 0.61 & 10.20 \\ \textbf{0.8} & \textbf{Time - PNS (sec)} & $<$0.01 & $<$0.01 & 0.07 & 1.14 & 213.31 \\ \textbf{0.8} & \textbf{Goodness Rate} & 0.97 & $>$0.99 & $>$0.99 & $>$0.99 & $>$0.99 \\
\hline
\textbf{1.0} & \textbf{Time - Google OR (sec)} & $<$0.01 & $<$0.01 & 0.03 & 0.29 & 4.01 \\ \textbf{1.0} & \textbf{Time - PNS (sec)} & $<$0.01 & $<$0.01 & 0.07 & 0.91 & 22.60 \\ \textbf{1.0} & \textbf{Goodness Rate} & 1.00 & 1.00 & 1.00 & 1.00 & 1.00 \\
\hline
\end{tabular}
}
\end{center}
\scriptsize
\end{table*}

The experimental results for the ratio of the transactions leaked to malicious organizations with respect to all available transactions in the network are given in Table 3 in which the ratio varies from 0.2 to 1.0. The first point to consider in the table is that the goodness rates perform better with increasing transaction leakage ratio. From the perspective of the malicious organization, the least ideal scenario is when no transactions are leaked (i.e. transaction leakage ratio is zero) since it is not possible to efficiently narrow down the user balance ranges. However, the users without any transactions are exceptions in that scenario since their balances can be still predictable as zero. On the other hand, the most ideal scenario is when all the transactions are leaked (i.e. transaction leakage ratio is one). The second point to consider is that the the times required to solve the minimum cost flow networks decrease with increasing transaction leakage ratio. The  reason behind this behaviour may be that as more  transactions are leaked, the greater the number of edges will be
in the network that have the same lower and upper bounds. Some minimum cost flow algorithms have capacity term in their complexity formula~\cite{ahuja1988} and hence consequently, the network becomes less complex to be solved.

\section{Conclusion}

Privacy is currently among the challenging issues in the blockchain that can be addressed from different perspectives. In the first part of this paper, we have considered private transaction problem on Ethereum blockchain where user balances and transaction amounts are shielded while the user addresses are still publicly traceable. For the given problem, a five-staged  private token transfer system has been proposed by integrating the zero-knowledge proof protocol into the private token smart contract. In the first stage, the token is deployed on the blockchain. In the second and third stages, two-way handshake is required between the parties of the transaction while in the fourth and the fifth stages, the real token transmission starts by constructing sender and receiver proofs. In the second part of the paper, we have made security analysis for the proposed transfer system by considering two different attacks as replay attack and balance range privacy attack which is modelled as a minimum cost flow network. 

For the experiments of the first part, we have considered the gas consumption of the smart contract where the functions corresponding to the private depositing and private withdrawing consumes the highest gas after the contract deployment. The proof generation times inside the browser may take up over two minutes. For the experiments of the second part, the network solution times and the goodness rates are provided with respect to number of addresses, number of transactions and transaction leakage ratio. The experimental results show the validity of our approach for the balance range privacy attack where it may be possible to estimate the possible ranges of user balances in case some transactions are deliberately leaked/sold to the particular organizations in blockchain.

\end{document}